\documentclass[sigplan]{acmart}

\usepackage{subcaption}
\usepackage{graphicx}
\usepackage{amsfonts}
\usepackage{multirow}
\usepackage{url}
\usepackage{listings}
\usepackage{tikz}
\usetikzlibrary{matrix}
\usetikzlibrary{positioning}
\usetikzlibrary{shapes.geometric}
\tikzstyle{every node}=[font=\footnotesize]
\tikzstyle{etichetta}=[]
\tikzstyle{square}=[draw,outer sep=0pt,inner sep=-1.3em,regular polygon,regular polygon sides=4,,minimum size=13.5mm]
\tikzstyle{square2}=[square,dotted,black!50]
\pgfmathsetmacro{\ra}{0.95}
\usetikzlibrary{decorations.pathreplacing,calc}

\usepackage{listings}
\lstdefinelanguage{json}{
    showspaces=false,
    showtabs=false,
    breaklines=false,
    breakatwhitespace=false,
    basicstyle=\ttfamily\footnotesize,
    upquote=true,
    morestring=[b]",
    literate={♭}{$\flat$}1
}
\usepackage{algorithm}
\usepackage{algpseudocode}

\algnewcommand\algorithmicswitch{\textbf{switch}}
\algnewcommand\algorithmiccase{\textbf{case}}
\algnewcommand\algorithmicdefault{\textbf{default}}
\algnewcommand\algorithmicassert{\texttt{assert}}
\algnewcommand\Assert[1]{\State \algorithmicassert(#1)}%
\algdef{SE}[SWITCH]{Switch}{EndSwitch}[1]{\algorithmicswitch\ #1\ \algorithmicdo}{\algorithmicend\ \algorithmicswitch}%
\algdef{SE}[CASE]{Case}{EndCase}[1]{\algorithmiccase\ #1}{\algorithmicend\ \algorithmiccase}%
\algdef{SE}[DEFAULT]{Default}{EndCase}[1]{\algorithmicdefault\ #1}{\algorithmicend\ \algorithmiccase}%
\algtext*{EndSwitch}%
\algtext*{EndCase}%
\algnewcommand\algorithmicforeach{\textbf{for each}}
\algdef{S}[FOR]{ForEach}[1]{\algorithmicforeach\ #1\ \algorithmicdo}

\renewcommand\footnotetextcopyrightpermission[1]{}
\acmConference[]{}{}

\begin{document}

\title{
Decoupling Trust in Byzantine CRDTs: \\ Fine-grained Post-Compromise Handling without Breaking Causality}

\author{Amos Brocco}
\email{amos.brocco@supsi.ch}
\orcid{0000-0002-0262-2044}
\affiliation{%
  \institution{University of Applied Sciences and Arts of Southern Switzerland}
  \city{Lugano}
  \country{Switzerland}
}

\begin{abstract}
Conflict-free Replicated Data Types (CRDTs) provide strong eventual consistency without coordination, but classical approaches assume benign participants. In Byzantine settings, convergence is typically enforced through agreement on update validity, often relying on identity-based filtering. However, such approaches struggle in post-compromise scenarios, where a previously correct participant becomes malicious: retroactive exclusion of its updates may break causal dependencies and invalidate subsequent computations.
In this paper, we decouple identity-based trust from content-based trust and introduce a fine-grained trust model that combines both dimensions. Building on deterministic reconstruction, our approach allows replicas to preserve previously accepted updates while enabling selective inclusion or exclusion based on both the originating identity (e.g., public keys) and the semantics of individual updates. Trust decisions can incorporate application-level policies, enabling precise control over the impact of each update on the system state.
Our approach preserves causal consistency and enables robust and flexible handling of both Byzantine and faulty behavior in decentralized CRDT systems.
\end{abstract}
\maketitle

\section{Introduction}

Byzantine behavior in distributed systems can lead to unpredictable and inconsistent outcomes, making resilience to adversarial participants a fundamental requirement for reliable system design. Such behavior may arise from both malicious intent (e.g., nodes attempting to violate system rules or disrupt operation) and accidental faults (e.g., software bugs causing arbitrary data generation or communication anomalies).

In this work, we focus on a specific class of distributed systems concerned with data storage and replication, namely Conflict-free Replicated Data Types (CRDTs)~\cite{b1,b10,b32}. CRDTs enable Strong Eventual Consistency (SEC) without coordination, allowing replicas to evolve independently while guaranteeing convergence when they observe the same set of updates~\cite{b0}. They span a wide range of data structures, from simple types such as counters, registers, and sets~\cite{b10} to more complex replicated objects~\cite{b4,b13}, and are typically classified into operation-based, state-based, and delta-state variants. While operation-based CRDTs~\cite{b2} propagate individual updates, state-based CRDTs~\cite{b10} exchange full states, simplifying correctness reasoning at the cost of higher communication overhead. Delta-state CRDTs ($\delta$-CRDTs)~\cite{b3,b14} strike a balance by only disseminating compact, idempotent deltas.

Classical CRDT designs assume benign participants, but in open and decentralized settings nodes may exhibit Byzantine behavior by generating malformed, conflicting, or malicious updates. As shown by Kleppmann~\cite{b23}, convergence is not guaranteed in such scenarios without additional mechanisms, typically based on deterministic validation of updates derived from their causal dependencies. Most existing approaches address this issue by revoking trust in participant identities, typically through public keys, and filtering their subsequent communications.

However, this strategy presents a fundamental limitation in CRDT-based systems due to causality. Updates are inherently interdependent: each modification generally builds upon previous ones. Consequently, retroactively excluding an update produced by a malicious node can invalidate subsequent updates that depend on it, even if those were generated by correct participants. This challenge is further exacerbated by the fact that Byzantine behavior is often not observable from the outset. In practice, a participant may behave correctly for an extended period before becoming compromised — for instance due to private key leakage — and subsequently appear to be acting maliciously. Revoking trust in such cases risks discarding previously correct contributions.

In systems with a trusted global notion of time, one might attempt to ignore updates produced after a given point (e.g., once a participant is deemed Byzantine). However, such an approach is not viable in decentralized, offline-first, eventually consistent environments, where timestamps are neither globally synchronized nor trustworthy, and may be forged to retroactively inject malicious updates.

A natural alternative is to rely on the causal structure of updates rather than explicit time. Byzantine-resilient CRDTs often use DAG-like structures built from cryptographic hashes of update content to ensure replica convergence~\cite{b23,b27}. These structures capture causal dependencies and define a partial order over updates. However, while this ordering reflects how updates relate to one another, it does not provide sufficient information to determine which updates should be considered before or after the point at which a participant is no longer trusted.

In particular, updates generated by a participant while offline may be disclosed only at a later time, potentially interleaving with newer updates from other replicas. As a result, neither timestamps nor causal ordering alone allow replicas to reliably distinguish past and future behavior, or to determine from which point revocation should be applied.

To address these limitations, we propose a practical approach for handling post-compromise behavior without breaking causality. The key idea is to decouple identity-based trust from content-based trust, allowing previously accepted updates to remain valid while preventing future malicious contributions from affecting the system. This separation ensures that Byzantine-produced data can be rendered harmless without disrupting the causal structure underlying CRDTs.

\section{Related Work}
\label{sec:related}

The problem of handling Byzantine behavior in CRDT-based systems is commonly addressed by controlling which updates are allowed to affect the replicated state. As summarized in Figure~\ref{fig:comparison}, existing approaches differ primarily in how they exclude, constrain, or reinterpret updates produced by adversarial participants.

\begin{figure*}[t]
\centering
\small
\begin{tabular}{lcc}
\textbf{Approach} & \textbf{Strategy} & \textbf{Byzantine Handling} \\
\hline
Validation (Kleppmann) \cite{b23} & Accept / reject & Filter by predicates \\
Proof-carrying CRDTs \cite{b34} & Accept if proof holds & Cryptographic validation \\
Equivocation-tolerant (Jacob) \cite{b24,b28} & Accept as concurrent & Absorb equivocation \\
Blocklace \cite{b27} & Validate + detect & Exclude faulty identities \\
SRDT \cite{b26} & Centralized filtering & Restrict via policy \\
Bounded CRDT (Baquero et al.) \cite{b33} & Cost bounding & Limit contribution rate \\
Deterministic reconstruction (Melda) \cite{b35} & Accept all / deterministic projection & Ignore via reconstruction \\
\textbf{This work} & Content-aware trust & Fine-grained trust-based filtering \\
\end{tabular}
\caption{Comparison of approaches for handling Byzantine updates in CRDT-based systems.}
\label{fig:comparison}
\end{figure*}

A fundamental challenge in Byzantine CRDTs is that convergence is no longer guaranteed when replicas make inconsistent decisions about which updates to accept. Kleppmann~\cite{b23} shows that divergence arises precisely when replicas disagree on update validity. To address this issue, validation-based approaches enforce deterministic predicates on updates derived from their causal dependencies, ensuring convergence by requiring that all correct replicas agree on which updates are admissible.

Recent work has refined this idea by reducing the cost of validation rather than its role. Proof-carrying CRDTs~\cite{b34} attach succinct cryptographic proofs to each update, attesting that the update and its causal history satisfy a predefined validity predicate. This enables non-interactive validation without access to the full history, but preserves the same underlying mechanism: convergence relies on consistent validation decisions across replicas.

An alternative perspective is provided by Jacob et al.~\cite{b24}, who introduce the notion of \textit{equivocation tolerance}. In this model, Byzantine updates are not rejected but incorporated as concurrent operations, and therefore do not violate convergence. This idea is further generalized using extend-only directed posets (EDPs)~\cite{b28}, where convergence follows from monotonic growth and deterministic merging. While these approaches eliminate the need for agreement on update admissibility, they do not limit the accumulation of adversarial updates.

DAG-based CRDT designs adopt this model by organizing updates as causally ordered, hash-linked structures. In such systems, Byzantine updates can be incorporated without violating convergence, but may accumulate unrestrained and affect system behavior. This motivates approaches that combine causal structures with mechanisms for detecting or constraining malicious activity.

Blocklace~\cite{b27} represents one such approach, combining DAG-based replication with identity-based validation. Updates are signed and verified, enabling the detection of equivocation and the eventual exclusion of Byzantine participants. Blocklace ensures that Byzantine behavior affects only a finite prefix of the computation through identity-based exclusion. However, this approach does not provide fine-grained control over which updates are retained or discarded, as exclusion is tied to identities rather than individual updates and is applied uniformly once a participant is deemed Byzantine. In contrast, our model enables content-based trust decisions, allowing selective preservation of past updates and more precise handling of post-compromise behavior.

Similarly, Secure RDTs (SRDTs)~\cite{b26} rely on centralized policy enforcement to filter updates. SRDTs avoid Byzantine inconsistencies by assuming a trusted central authority and a static security policy, whereas our work addresses trust evolution in fully decentralized settings. Bounded Byzantine CRDTs~\cite{b33} instead limit adversarial impact by associating a computational cost to updates, thereby constraining the rate at which Byzantine nodes can inject updates. However, they do not distinguish between benign and malicious updates, nor do they provide mechanisms for selectively retaining or discarding updates based on their content. Our approach is orthogonal, focusing instead on selectively interpreting updates based on trust and application-level semantics.

Our previous work on deterministic reconstruction~\cite{b35} introduces a different design point. Rather than deciding which updates are admissible, all updates are propagated and stored, while only a deterministically defined subset contributes to the reconstructed state. In this model, convergence does not depend on agreement on update validity, but on deterministic interpretation of a shared set of updates. Byzantine behavior is handled by ensuring that adversarial updates are either structurally rejected or deterministically ignored during reconstruction, without requiring global coordination or validation agreement.

Despite these differences, most existing approaches implicitly assume that Byzantine behavior must be handled by controlling the set of updates that affect the system, either through validation, exclusion, or resource constraints. However, this assumption becomes problematic in the presence of \textit{post-compromise Byzantine behavior}. In practical systems, a node may behave correctly for a period of time before becoming compromised, for example due to key leakage, and subsequently acting maliciously. In such cases, identity-based exclusion or validation-based filtering may retroactively invalidate updates that were correct when produced.

In causally structured CRDTs, this problem is particularly acute. Updates are interdependent, and excluding a past update may invalidate all subsequent updates that depend on it, even if those were generated by correct participants. As a result, exclusion is not a local operation but can have cascading effects on the causal history.

In systems with a trusted notion of time, one could attempt to mitigate this issue by ignoring updates produced after a given point. However, in decentralized, offline-first settings, timestamps are neither globally synchronized nor trustworthy, and cannot be used to reliably separate benign from malicious updates.

This work builds upon deterministic reconstruction~\cite{b35} and extends it with explicit support for trust evolution. We shift the focus from preventing or validating Byzantine updates to controlling their effects once they occur. By decoupling identity-based trust from content-based trust, we introduce an exclusion model that can preserve previously accepted updates while preventing future malicious contributions. This enables post-compromise handling without breaking causal dependencies.

\section{Melda}
\label{sec:melda}

We briefly recall the key concepts of Melda\footnote{\url{https://github.com/slashdotted/libmelda}}, a delta-state CRDT for JSON documents, focusing on the aspects relevant to Byzantine resilience and trust management. A complete description of the data structure and its design is provided in~\cite{b20,b22}, while formal proofs of Byzantine resilience are presented in~\cite{b35}.

\subsection{System model}

Melda is a delta-state replication framework operating in a decentralized, asynchronous, and offline-first setting. Replicas generate local updates encoded as immutable \textit{delta blocks}, which are exchanged and accumulated in a grow-only structure. Each delta block references a set of prior blocks, called \textit{anchors}, representing its causal dependencies.

Delta blocks are identified by a cryptographic hash and form a hash-linked directed acyclic graph (DAG). A monotonic index ensures acyclicity and defines a partial causal order. Validation and application of updates depend only on the content of the block and its declared dependencies, and are therefore deterministic across replicas that observe the same data.

\subsection{Data model and reconstruction}

Melda operates directly on the application's JSON data model by decomposing it into a collection of immutable objects and revisions. Updates generate new revisions, which are organized into per-object revision trees.
The application state is obtained through a deterministic reconstruction process that resolves references starting from a root object and selects a \textit{winning revision} for each object. Only revisions that are causally reachable from the root contribute to the reconstructed state, while unrelated or inconsistent branches are deterministically ignored.
This reconstruction model ensures that the resulting state depends only on the set of available updates, not on their order of application.

\subsection{Validation and applicability}

Upon reception, delta blocks are validated locally by checking structural correctness, hash integrity, and the availability of their dependencies. A block is applicable if all its anchors have already been applied.
Both validation and applicability are deterministic functions of the block content and its dependencies. As a result, replicas that observe the same set of updates make identical decisions about which updates are accepted and when they are applied.

\section{Deterministic reconstruction and Byzantine resilience}

A key property of Melda is that convergence is ensured by deterministic state reconstruction rather than by agreement on update validity. As shown in~\cite{b35}, if two replicas observe the same set of delta blocks, they deterministically derive the same state, even in the presence of Byzantine behavior such as equivocation, omission, or arbitrary update injection.
This model decouples update propagation from state derivation: all updates may be disseminated and stored, but only a deterministically defined subset contributes to the reconstructed state.

\subsection{Security extension: \textit{melda-sec}}
\label{sec:melda-sec}

Melda can be extended with a pluggable security layer, \textit{melda-sec}\footnote{\url{https://github.com/slashdotted/libmelda-sec}}, implemented as an adapter that composes with the underlying storage and transport mechanisms.
Each delta block may be signed, and replicas maintain a local set of trusted public keys. The validation predicate is refined to include signature verification and policy checks. At reconstruction time, only updates satisfying these constraints contribute to the state, while others are ignored. Importantly, all updates are still propagated and stored, independently of their trust status. This preserves availability and allows replicas to apply different trust policies over the same underlying set of updates. Replicas with identical trust configurations converge to the same state, while different configurations may yield different but internally consistent views.

\section{Trust evolution and post-compromise handling}

\begin{figure}[t]
    \centering
    \includegraphics[width=.7\linewidth]{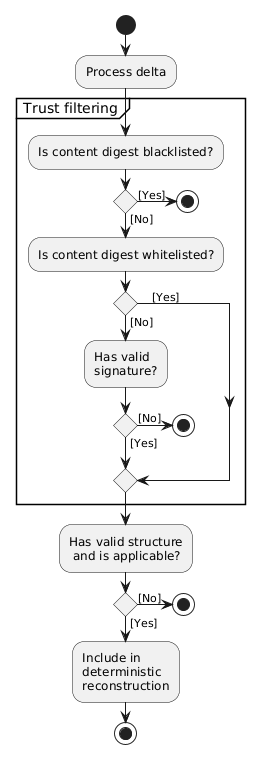}
    \caption{Fine-grained trust filtering applied to each delta block. Each update is first checked against a content-based blacklist, then a whitelist, and, if no explicit decision is made, validated through signature verification before deterministic reconstruction.}
    \label{fig:trust-flow}
\end{figure}

In this work, the trust model provided by \textit{melda-sec} is extended to address a fundamental limitation of identity-based approaches: trust is not static over time. In practical systems, a replica may behave correctly for a long period before becoming compromised, for instance due to private key leakage. In such situations, updates produced before the compromise should remain valid, while subsequent updates should be treated as untrusted.

Traditional approaches based on identity revocation are not well suited to capture this distinction. Once an identity is deemed Byzantine, its contributions are typically handled uniformly, without distinguishing between individual updates. This is the case in identity-based exclusion mechanisms such as Blocklace~\cite{b27}, where updates from a participant are eventually disregarded once it is classified as faulty.

While effective at limiting the long-term influence of Byzantine nodes, this approach lacks the ability to perform fine-grained exclusion. In particular, it cannot selectively retain updates that remain valid or semantically meaningful while discarding only those that are undesirable.

This limitation is particularly problematic in causally structured systems, where updates are interdependent. Since identity-based decisions do not differentiate between individual contributions, they may prevent the inclusion of updates that would otherwise remain valid, together with the updates that depend on them. As a result, the application cannot preserve a consistent subset of the causal history when trust in a participant changes.

To address this limitation, we build on the content-based filtering mechanisms of \textit{melda-sec} and introduce a trust model that decouples identity-based trust from content-based trust. Rather than associating trust decisions solely with identities, our approach applies them at the level of individual delta blocks, enabling fine-grained control over which updates contribute to the system state. Figure~\ref{fig:trust-flow} illustrates this process, where each delta block is first checked against a content-based blacklist, then against a whitelist, and, if not explicitly accepted, validated through signature verification. This contrasts with identity-based approaches, where exclusion is applied uniformly once a participant is deemed faulty.

More concretely:
\begin{itemize}
    \item public keys determine whether newly received updates are considered for validation, enabling selective identity-based trust or revocation;
    \item content-based whitelist mechanisms explicitly preserve previously accepted delta blocks, even if their origin subsequently becomes untrusted;
    \item content-based blacklist mechanisms enable the selective exclusion of specific updates without affecting unrelated portions of the causal history.
\end{itemize}

Importantly, whitelist and blacklist decisions can be derived from application-level validation and control logic. While the CRDT layer enforces structural correctness—such as well-formed data, valid causal references, and correct signatures—it does not capture application-specific notions of correctness or benign behavior. As a result, updates that satisfy all formal validity conditions may still be undesirable. By expressing trust at the content level, the system enables applications to refine acceptance criteria beyond intrinsic CRDT validity.

This design enables precise handling of post-compromise behavior: updates produced before a compromise can be preserved, while future contributions from the same identity can be selectively excluded.

\paragraph{Example (whitelisting).}
Consider a distributed application for managing orders, where each replica maintains a local data model representing orders and their state (e.g., \textit{pending}, \textit{confirmed}, \textit{shipped}). The application serializes this state into a JSON document which is replicated using Melda as a $\delta$-CRDT.
Assume that a participant $p$ behaves correctly for some time, producing valid updates that are incorporated into the system and may be referenced by other operations. At a later point, $p$ becomes compromised and starts issuing malicious updates, for instance by altering the status of existing orders or injecting unauthorized changes into the JSON representation.
In identity-based exclusion approaches, once $p$ is deemed Byzantine, all updates originating from $p$ are excluded, including those that were previously valid and contributed to the construction of the order objects. Since other updates may depend on these earlier modifications, their exclusion can invalidate parts of the order history, potentially breaking the causal structure and leading to inconsistent or incomplete application state.
In contrast, our approach enables fine-grained trust decisions at the level of individual updates. Valid updates produced by $p$ prior to the compromise can be preserved through whitelisting, ensuring that previously established order data remains intact. At the same time, malicious updates—such as unauthorized status transitions in the JSON data—can be selectively excluded through blacklist mechanisms.

\paragraph{Example (blacklisting).}
The same mechanism can also be applied in non-malicious scenarios. For example, suppose that a participant operates a buggy version of the application that generates incorrect updates for a subset of operations, such as invalid status transitions or malformed field values. Instead of excluding all updates from that participant, replicas can selectively blacklist only the affected delta blocks once the fault is identified, while preserving all other valid contributions. This allows the system to correct localized inconsistencies without discarding unaffected parts of the history, highlighting that fine-grained trust mechanisms are useful not only for handling adversarial behavior, but also for isolating and correcting non-malicious faults.
\vspace{1em}

\subsection{Discussion}
The previous examples illustrate how the proposed model enables fine-grained trust decisions in practice. We now discuss the broader implications of this design.

In the proposed trust model, filtering decisions can be driven by application-level semantics of the data model, rather than by structural validity alone. While Melda ensures that all updates are well-formed and causally consistent, the application can interpret the JSON representation to detect and filter semantically invalid changes. This allows the system to retain a consistent and causally valid history while preventing further influence from compromised or faulty participants.

These examples show that trust decisions cannot be expressed solely at the level of identities or global policies, but must instead operate at the granularity of individual updates. They also demonstrate that correctly handling compromise requires distinguishing between prior and subsequent behavior.

This observation raises a key question: how can this distinction be expressed in a decentralized setting, without breaking causal dependencies or discarding valid historical updates? In particular, once trust in a participant changes, replicas must determine from which point its contributions should no longer be considered reliable, without compromising the integrity of the causal history.

One possible approach is to rely on temporal information, by introducing a cutoff that separates updates produced before and after a given point in time. However, such an approach is not viable in decentralized CRDT systems. In the absence of a trusted global clock, timestamps are neither globally synchronized nor tamper-proof, and can be arbitrarily manipulated by Byzantine participants. Furthermore, in offline-first environments, updates may be created and disseminated out of order, making it impossible to reliably associate them with a global timeline.

As a result, trust evolution cannot be reliably expressed in terms of time alone, and must instead be defined over the causally observed set of updates. Our model addresses this challenge by shifting trust decisions from time-based reasoning to causality-aware filtering, allowing replicas to preserve previously accepted updates while selectively filtering subsequent ones according to updated trust policies, without requiring retroactive invalidation.

This approach integrates naturally with deterministic reconstruction, ensuring that filtering decisions remain consistent with the underlying causal structure. Since filtering is applied at reconstruction time and depends only on the set of updates and the local trust configuration, replicas sharing the same configuration converge to the same state. Different configurations may yield different views, but each remains internally consistent.

Overall, this model shifts the focus from identity revocation to fine-grained, content-aware, and causality-aware trust management, enabling robust handling of compromise without breaking causal dependencies or requiring coordination.
We note that the distribution and agreement on trust policies is orthogonal to our model. Our approach defines the semantics of state reconstruction given a particular trust configuration, but does not impose how such configurations are established or synchronized across replicas. Mechanisms for sharing or agreeing on trust policies, such as quorum-based endorsement, are left as future work.

\section{Conclusion}
\label{sec:conclusion}
In this paper, we revisited the problem of Byzantine behavior in CRDT-based systems through the lens of post-compromise scenarios. While existing approaches primarily focus on preventing or validating malicious updates, they implicitly assume that trust in identities can be revoked without affecting previously accepted updates. We showed that this assumption does not hold in causally structured systems, where updates are interdependent and retroactive exclusion can break the integrity of the computation.

Building on deterministic reconstruction, we introduce a model that decouples identity-based trust from content-based trust and enables fine-grained trust evolution. In our approach, trust decisions are applied before reconstruction, allowing previously accepted updates to be preserved while selectively filtering subsequent contributions from compromised identities.

A key aspect of our proposal is that trust is expressed at the level of individual updates and can incorporate application-specific validation logic. This allows the system to go beyond structural correctness and control the semantic impact of each update, while maintaining deterministic convergence within each trust configuration. The resulting model preserves causal consistency without requiring global coordination or trusted timestamps.

Overall, our work shifts the focus from identity revocation and update admissibility to fine-grained, causality-aware trust management. By treating trust as a dynamic property applied to individual updates rather than static identities, we enable robust and flexible handling of compromise in decentralized, offline-first environments.

Future work includes exploring automated strategies for trust adaptation, investigating interoperability between different trust configurations, and integrating lightweight mechanisms to strengthen guarantees on update agreement across replicas.

In particular, an interesting direction is the use of quorum-based endorsement, where replicas attach digital signatures to delta blocks to indicate acceptance. In a permissioned setting with known participants, such endorsements can be disseminated alongside updates — for example, as auxiliary signatures associated with a given delta. A delta may then be considered trusted only after receiving a sufficient number of endorsements, enabling replicas to converge not only on state derivation but also on a shared set of accepted updates.

This approach is orthogonal to our model and complements deterministic reconstruction by providing additional guarantees on update agreement, while preserving the decentralized and causality-aware nature of the system.

\bibliographystyle{acm}
\bibliography{melda-bft}

@InBook{	  b0,
  title		= {{Conflict-Free Replicated Data Types CRDTs}},
  author	= {Pregui{\c{c}}a, Nuno and Baquero, Carlos and Shapiro,
		  Marc},
  year		= 2018,
  booktitle	= {Encyclopedia of Big Data Technologies},
  publisher	= {Springer International Publishing},
  address	= {Cham},
  pages		= {1--10},
  isbn		= {978-3-319-63962-8}
}

@Article{	  b1,
  title		= {{Consistency without Concurrency Control in Large, Dynamic
		  Systems}},
  author	= {Letia, Mihai and Pregui\c{c}a, Nuno and Shapiro, Marc},
  year		= 2010,
  month		= apr,
  journal	= {SIGOPS Oper. Syst. Rev.},
  publisher	= {Association for Computing Machinery},
  address	= {New York, NY, USA},
  volume	= 44,
  number	= 2,
  pages		= {29–34},
  issn		= {0163-5980},
  issue_date	= {April 2010},
  numpages	= 6
}

@InProceedings{	  b2,
  title		= {{Making Operation-Based CRDTs Operation-Based}},
  author	= {Baquero, Carlos and Almeida, Paulo S\'{e}rgio and Shoker,
		  Ali},
  year		= 2014,
  booktitle	= {Proceedings of the First Workshop on Principles and
		  Practice of Eventual Consistency},
  location	= {Amsterdam, The Netherlands},
  publisher	= {Association for Computing Machinery},
  address	= {New York, NY, USA},
  series	= {PaPEC '14},
  isbn		= 9781450327169,
  articleno	= 7,
  numpages	= 2
}

@Article{	  b3,
  title		= {{Delta state replicated data types}},
  author	= {Paulo S\'{e}rgio Almeida and Ali Shoker and Carlos
		  Baquero},
  year		= 2018,
  journal	= {Journal of Parallel and Distributed Computing},
  volume	= 111,
  pages		= {162--173},
  issn		= {0743-7315}
}

@Article{	  b4,
  title		= {{A Conflict-Free Replicated JSON Datatype}},
  author	= {Kleppmann, Martin and Beresford, Alastair R.},
  year		= 2017,
  journal	= {IEEE Transactions on Parallel and Distributed Systems},
  volume	= 28,
  number	= 10,
  pages		= {2733--2746}
}

@TechReport{	  b10,
  title		= {{A comprehensive study of Convergent and Commutative
		  Replicated Data Types}},
  author	= {Shapiro, Marc and Pregui{\c c}a, Nuno and Baquero, Carlos
		  and Zawirski, Marek},
  year		= 2011,
  month		= jan,
  number	= {RR-7506},
  pages		= 50,
  type		= {Research Report},
  institution	= {Inria -- Centre Paris-Rocquencourt ; INRIA}
}

@Misc{		  b13,
  title		= {{Yjs: A Framework for Near Real-Time P2P Shared Editing on
		  Arbitrary Data Types}},
  author	= {Nicolaescu, Petru and Jahns, Kevin and Derntl, Michael and
		  Klamma, Ralf},
  year		= 2015,
  month		= {06},
  isbn		= {978-3-319-19889-7}
}

@InProceedings{	  b14,
  title		= {{Array CRDTs Using Delta-Mutations}},
  author	= {Rinberg, Arik and Solomon, Tomer and Khazma, Guy and
		  Lushi, Gal and Shlomo, Roee and Ta-Shma, Paula},
  year		= 2021,
  month		= apr,
  booktitle	= {8th Workshop on Principles and Practice of Consistency for
		  Distributed Data},
  publisher	= {ACM},
  series	= {PaPoC 2021},
  articleno	= 3
}

@inproceedings{b20,
author = {Brocco, Amos},
title = {Melda: A General Purpose Delta State JSON CRDT},
year = {2022},
isbn = {9781450392563},
publisher = {Association for Computing Machinery},
address = {New York, NY, USA},
url = {https://doi.org/10.1145/3517209.3524039},
doi = {10.1145/3517209.3524039},
booktitle = {Proceedings of the 9th Workshop on Principles and Practice of Consistency for Distributed Data},
pages = {1â€“7},
numpages = {7},
location = {Rennes, France},
series = {PaPoC '22}
}

@misc{b22,
      title={Introducing Support for Move Operations in Melda CRDT}, 
      author={Amos Brocco},
      year={2025},
      eprint={2503.04811},
      archivePrefix={arXiv},
      primaryClass={cs.PL},
      url={https://arxiv.org/abs/2503.04811}, 
}

@inproceedings{b23,
author = {Kleppmann, Martin},
title = {Making CRDTs Byzantine fault tolerant},
year = {2022},
isbn = {9781450392563},
publisher = {Association for Computing Machinery},
address = {New York, NY, USA},
url = {https://doi.org/10.1145/3517209.3524042},
doi = {10.1145/3517209.3524042},
booktitle = {Proceedings of the 9th Workshop on Principles and Practice of Consistency for Distributed Data},
pages = {8–15},
numpages = {8},
location = {Rennes, France},
series = {PaPoC '22}
}

@misc{b24,
author = "Jacob, Florian and Bayreuther, Saskia and Hartenstein, Hannes",
title = "On CRDTs in Byzantine Environments",
year = 2022,
doi = "10.18420/sicherheit2022_07",
howpublished = "GI SICHERHEIT 2022",
publisher = "Gesellschaft für Informatik, Bonn",
pissn = "1617-5468",
isbn = "978-3-88579-717-3",
pages = "113--126",
}

@article{b26,
author = {Renaux, Thierry and Van den Vonder, Sam and De Meuter, Wolfgang},
title = {Secure RDTs: Enforcing Access Control Policies for Offline Available JSON Data},
year = {2023},
issue_date = {October 2023},
publisher = {Association for Computing Machinery},
address = {New York, NY, USA},
volume = {7},
number = {OOPSLA2},
url = {https://doi.org/10.1145/3622802},
doi = {10.1145/3622802},
journal = {Proc. ACM Program. Lang.},
month = oct,
articleno = {227},
numpages = {27}
}

@misc{b27,
      title={The Blocklace: A Byzantine-repelling and Universal Conflict-free Replicated Data Type}, 
      author={Paulo Sérgio Almeida and Ehud Shapiro},
      year={2025},
      eprint={2402.08068},
      archivePrefix={arXiv},
      primaryClass={cs.DC},
      url={https://arxiv.org/abs/2402.08068}, 
}

@inproceedings{b28,
author = {Jacob, Florian and Hartenstein, Hannes},
title = {On Extend-Only Directed Posets and Derived Byzantine-Tolerant Replicated Data Types},
year = {2023},
isbn = {9798400700866},
publisher = {Association for Computing Machinery},
address = {New York, NY, USA},
url = {https://doi.org/10.1145/3578358.3591333},
doi = {10.1145/3578358.3591333},
booktitle = {Proceedings of the 10th Workshop on Principles and Practice of Consistency for Distributed Data},
pages = {63–69},
numpages = {7},
location = {Rome, Italy},
series = {PaPoC '23}
}

@inproceedings{b32,
author = {Shapiro, Marc and Pregui\c{c}a, Nuno and Baquero, Carlos and Zawirski, Marek},
title = {Conflict-free replicated data types},
year = {2011},
isbn = {9783642245497},
publisher = {Springer-Verlag},
address = {Berlin, Heidelberg},
booktitle = {Proceedings of the 13th International Conference on Stabilization, Safety, and Security of Distributed Systems},
pages = {386–400},
numpages = {15},
location = {Grenoble, France},
series = {SSS'11}
}

@inproceedings{b33,
author = {Baquero, Carlos and Maia, Francisco and Dantas, Abel and Anta, Antonio Fern\'{a}ndez and Frey, Davide and S\'{a}nchez, C\'{e}sar and Albouy, Timoth\'{e}},
title = {Bounding Byzantine Impact in Open CRDT Systems},
year = {2026},
isbn = {9798400726378},
publisher = {Association for Computing Machinery},
address = {New York, NY, USA},
url = {https://doi.org/10.1145/3806077.3806698},
doi = {10.1145/3806077.3806698},
booktitle = {Proceedings of the 13th International Workshop on Principles and Practice of Consistency for Distributed Data},
pages = {17–24},
numpages = {8},
location = {
},
series = {PaPoC '26}
}

@inproceedings{b34,
author = {Marx, Nick and Jacob, Florian and Hartenstein, Hannes},
title = {Proof-Carrying CRDTs allow Succinct Non-Interactive Byzantine Update Validation},
year = {2025},
isbn = {9798400715587},
publisher = {Association for Computing Machinery},
address = {New York, NY, USA},
url = {https://doi.org/10.1145/3721473.3722142},
doi = {10.1145/3721473.3722142},
booktitle = {Proceedings of the 12th Workshop on Principles and Practice of Consistency for Distributed Data},
pages = {15–21},
numpages = {7},
location = {World Trade Center, Rotterdam, Netherlands},
series = {PaPoC '25}
}

@misc{b35,
      title={A Composable CRDT Layer for Byzantine-Resilient Deterministic Reconstruction}, 
      author={Amos Brocco},
      year={2026},
      eprint={2606.18966},
      archivePrefix={arXiv},
      primaryClass={cs.DC},
      url={https://arxiv.org/abs/2606.18966}, 
}
\end{document}